\documentclass{article}
\usepackage{spconf,amsmath,graphicx}

\usepackage{booktabs}
\usepackage{multirow}
\usepackage{subcaption}
\usepackage{cite}

\def\vertspace{\vspace{-0.4cm}}

\title{On Unsupervised Uncertainty-Driven Speech Pseudo-Label Filtering and Model Calibration}

\name{Nauman Dawalatabad$^1$, Sameer Khurana$^1$, Antoine Laurent$^2$, James Glass$^1$ \thanks{This work is supported by DSTA. A part of this work was performed using HPC resources by GENCI–IDRIS under allocation AD011012527.} }
\address{$^1$MIT Computer Science and Artificial Intelligence Laboratory, Cambridge, MA, USA\\ $^2$LIUM - Le Mans University, France}

\begin{document}
\ninept
\maketitle
\begin{abstract}
Pseudo-label (PL) filtering forms a crucial part of Self-Training (ST) methods for unsupervised domain adaptation. Dropout-based Uncertainty-driven Self-Training (DUST) proceeds by first training a teacher model on source domain labeled data. Then, the teacher model is used to provide PLs for the unlabeled target domain data. Finally, we train a student on augmented labeled and pseudo-labeled data. The process is iterative, where the student becomes the teacher for the next DUST iteration. A crucial step that precedes the student model training in each DUST iteration is filtering out noisy PLs that could lead the student model astray. In DUST, we proposed a simple, effective, and theoretically sound PL filtering strategy based on the teacher model's uncertainty about its predictions on unlabeled speech utterances. We estimate the model's uncertainty by computing disagreement amongst multiple samples drawn from the teacher model during inference by injecting noise via dropout. In this work, we show that DUST's PL filtering, as initially used, may fail under severe source and target domain mismatch. We suggest several approaches to eliminate or alleviate this issue. Further, we bring insights from the research in neural network model calibration to DUST and show that a well-calibrated model correlates strongly with a positive outcome of the DUST PL filtering step.

\end{abstract}
\section{Introduction}
\label{sec:intro}
In recent years, Automatic Speech Recognition (ASR) has improved dramatically due to the introduction of novel neural network architectures, training frameworks, and large labeled and unlabeled datasets~\cite{Graves2006,Graves2013,Chorowski2015,PoveyPGG16,Hori2017}. However, domain generalization remains an unsolved problem; an ASR model's performance drops significantly when the training and testing (or inference) conditions do not match. 

Several previous works attempt to address the issue of domain adaptation, such as consistency regularization, domain-adversarial training, multi-domain self-supervised pre-training, and the much simpler Self-Training (ST) method \cite{saito2018maximum,ganin2016domain,bell2020adaptation,review_domain_adapt21,he2019revisiting}. ST is useful when the target domain data distribution  differs significantly from the source domain data distribution (e.g., read speech to YouTube speech). ST proceeds by training a teacher model on a labeled set in the source domain, which generates pseudo-labels (PLs) for the unlabeled set  in the target domain. Then, a student model uses both labeled and pseudo-labeled data for training. We can iterate over the Teacher/Student training such that the student from a previous iteration acts as the teacher for the next ST iteration. Classical ST that uses all PLs might be sub-optimal as PLs might be noisy. To address this  issue, we previously proposed Dropout Uncertainty-driven Self-Training (DUST)~\cite{dust, 9746276}, which appends the original Self-Training algorithm with PL filtering step to weed out noisy PLs.

Intuitively, the PL filtering stage in DUST computes a proxy for the model's confidence in its predictions on an unlabeled speech utterance. It is computed using the agreement between a reference prediction and $T$ sampled predictions. If the model's confidence is below a pre-defined threshold,  we discard that utterance. The DUST filtering stage assumes that the model's confidence is a reliable estimate of PL quality, i.e., high confidence implies low Word Error Rate (WER) on accepted utterances. In this work, we found that the DUST algorithm fails when the domain mismatch is severe. Hence, the DUST filtering method could accept noisy PLs. 

This paper focuses on Pseudo-Label (PL) filtering stage in DUST. The goal of this work is to stress test  the DUST PL filtering under severe domain mismatch, and suggest  approaches to alleviate the DUST PL filtering issue.
The following are the main contributions to this work, 
(i) We show that the DUST algorithm fails when the domain mismatch is severe.
(ii) We propose several approaches to mitigate or eliminate this breakpoint.
(iii) We study, for the time, the PL filtering in the context of model calibration for DUST. 
(iv) We show that there is a strong correlation between the model calibration errors and the quality of the filtered PLs.

\section{Methodology}
\subsection{DUST based Pseudo-Label Filtering}

DUST is a method for domain adaptation of Automatic Speech Recognition (ASR) models. It proceeds by training a teacher on labeled source domain data, which is used to provide PLs on unlabeled target domain data. A student model is then trained on augmented source domain labeled and target domain pseudo-labeled data. Due to the source and target domain mismatch, the PLs could be highly erroneous. Hence, a crucial step in DUST is PL filtering. 

For each utterance in unlabelled target domain $x_u$, we sample  $T$ predictions $\{\hat{y}_{u}^{t_1}, \dots, \hat{y}_{u}^{t_T}\}$ by injecting dropout noise during inference and changing the random seed~\cite{gal2016dropout}. A reference hypothesis $\hat{y}_{u}^{ref}$ is also generated without dropout. DUST uses agreement between the reference and sampled predictions to measure the model’s predictive uncertainty as follows: 
\begin{equation}
pred\_uncert(x_u) =  \max_{t\in\{1,\dots, T\}}\{ eds_u^t \}    
\label{eq:uncert}
\end{equation}

\begin{equation}
eds_u^t =   editdistance( \hat{y}_{u}^{ref}, \hat{y}_{u}^{t})/|\hat{y}_u^{ref}| 
\label{eq:eds}
\end{equation}
 
 We exclude $(x_u, \hat{y}_{u}^{ref})$ with $pred\_uncert(x_u)>\tau$, where $\tau$ is a filtering threshold. Lower values of $\tau$ leads to less noisy PLs. 

\subsection{Approaches to Improve Pseudo-Label  Filtering in DUST} 
The low value of predictive uncertainty is a reliable indicator of PL quality. However, we observe that this assumption breaks under severe domain mismatch. We found that the DUST algorithm fails to filter good quality pseudo-labeled utterances in the initial stage of filtering (as demonstrated later in Section~\ref{sec:results}).
This is due to the model's high confidence (incorrectly) generated using the $T$ sampled predictions, where both the reference and the $T$ sampled predictions are decoded incorrectly.
We propose the following approaches to address this issue with DUST PL filtering algorithm.

\vertspace\paragraph*{Stricter consensus:}
The main step in the DUST PL filtering process is the consensus between the reference and the sampled predictions. Originally, DUST uses only three sampled predictions ($T$=3). However, three samples may not be enough to obtain robust estimates of predictive uncertainty on unlabeled points.  Hence, we propose to use more number of sampled predictions to compute the teacher model's predictive uncertainty on unlabeled target domain data. 
More number of sampled predictions leads to a stricter consensus policy as a large number of sampled predictions need to agree with the reference prediction such that all the $T$ edit distances between $\hat{y}_{u}^{ref}$ and $\{\hat{y}_{u}^{t}\}_{t=1}^T$ fall below a filtering threshold $\tau$.
This leads to robust estimates of the model's predictive uncertainty.

\vertspace\paragraph*{Modified uncertainty estimate:}
Another important factor is the choice of tokens for calculating the model's predictive uncertainty. As shown in Equation~(\ref{eq:eds}), the original DUST PL filtering algorithm measures the model's predictive uncertainty per reference-sample PL pair in terms of edit distance between the word sequences of $\hat{y}_{u}^{ref}$ and $\hat{y}_{u}^{t}$, i.e., $eds_u^t$.
However, the word-based edit distances might not be the optimal choice as it focuses on the coarse-grained structure of the candidate PL ($y_u^{ref}$) to be filtered. We propose a fine-grained version of DUST, Character DUST (C-DUST) that estimates the edit distance ($eds_u^t$) between the reference-sample PL pair $\hat{y}_{u}^{ref}$ and $\hat{y}_{u}^{t}$ using the character sequences as opposed to word-sequences. This helps the filtering algorithm to precisely select PLs by focusing on finer details.

For example, consider an utterance with the true label as $y_u^{true}$=``signs of ankylosing spondylitis detected'' for which model's predictions are; $\hat{y}_{u}^{ref}$=``signs of ankylosin spondylitis detected'', $\hat{y}_{u}^{t_1}$=``sgns o ankylosin spondylitis detectd'', and $\hat{y}_{u}^{t_2}$=``sgns of avkclozin sondilietis detected''. DUST can not distinguish between  $\hat{y}_{u}^{t_1}$ and $\hat{y}_{u}^{t_2}$ as both have $eds_{u}^{t_1}=eds_{u}^{t_2}=0.6$. Whereas, C-DUST can distinguish this using the finer tokens ($eds_{u}^{t_1}=0.085$ and $eds_{u}^{t_2}=0.2$) leading to precise uncertainty estimates. 

\vertspace\paragraph*{Robust teacher:}
The teacher model plays a vital role in PL filtering. In the previous DUST paper, we do not make any assumption about the teacher model's robustness and directly used the most robust model possible that already shows a reasonable performance on the target domain.  We then used DUST to further improve the performance on the target domain. This is a strong assumption and need not hold true in many cases, especially when domain mismatch is severe. 
We show improving the robustness of the teacher model plays a vital role in the DUST PL filtering algorithm which in turn helps alleviate the DUST PL filtering issue mentioned above.

\vertspace\paragraph*{Diverse source domain data:}
In the scenario where the source domain labeled data is available in abundance with diverse conditions (e.g., high-resource English datasets), it is intuitive to use more labeled source domain data for training the teacher model. This provides a better teacher model for PL filtering. Further, we demonstrate later (in Section \ref{sec:results}) that diverse source domain data along with data augmentation during teacher model training helps to filter out better target domain PLs.

\subsection{Model Calibration and Pseudo-Label filtering}
\label{sec:calibration}

Model calibration is a crucial step to bridge the gaps between better predictions and the true predictions. A well-calibrated model assigns prediction probability (i.e., confidence) close to the true correctness measure (i.e., accuracy) \cite{guo_calib:2017}. Let $h$ be a neural network model (teacher model).
Assuming $h(X) = (\hat{Y},\hat{C})$, where is $\hat{Y}$ and $\hat{C}$ are model's prediction and estimated confidence for the prediction. A perfect calibration is defined as follows~\cite{guo_calib:2017}:
\begin{equation}
    P(\hat{Y}=Y | \hat{C}=c) = c, \forall c \in [0,1]
\end{equation}

\vertspace\paragraph*{A measure of confidence in PL filtering:} In the context of PL filtering algorithm ($f$) with the teacher model ($h$), the overall model can be jointly defined as $f(h(x_u))$.
In order to measure how well the model is calibrated, we need to define \textit{accuracy} and \textit{confidence} for the overall PL filtering model. The PL filtering approach provides a proxy of confidence in selecting a PL.
It uses edit distance as a measure of the model's predictive uncertainty as given in Equation~(\ref{eq:uncert}). 

The label errors are a measure of dissimilarities. We define the confidence of the model for an utterance $x_u$ as: 
\begin{equation}
conf(x_u) = 1 - pred\_uncert(x_u) 
\label{eq:conf}
\end{equation}
Intuitively, edit distance can be thought of as a disagreement between a reference prediction $\hat{y}_u^{ref}$ and a sampled prediction $\hat{y}_{u}^{t}$. 
It denotes the model's uncertainty in predicting reference and sampled hypotheses. Overall, when all the hypotheses are similar, the model is less uncertain about its prediction and hence the edit distance will be low and confidence will be high. Similarly, we  define a model's true accuracy as follows:
\begin{equation}
    acc(x_u) = 1 -  err(x_u)
\end{equation}
where $err$ denotes true error rate on $x_u$. We lower bound $acc$ and $conf$ to zero. In this work, we use the following calibration error (CE) metrics~\cite{guo_calib:2017}:

\noindent\textbf{{Expected Calibration Error (ECE):-}}
The approximate Expected Calibration Error (ECE) is calculated by binning the set of confidence scores into $M$ disjoint bins as follows:
\begin{equation}
    ECE = \sum_{m=1}^{M} \frac{|B_m|}{n} |acc(B_m)-conf(B_m)|
    \label{eq:ece}
\end{equation}
where $n$ denotes the total number of utterances and $B_m$ denotes the $m$-th bin.
ECE is a weighted averages of calibration errors over the bins and denotes the overall miscalibration of the model. 
Another variant of ECE is the Root Mean Square ECE (RMS-ECE/RCE) that takes squares of difference of the scores instead of absolute differences~\cite{guo_calib:2017}. 

\noindent\textbf{{Maximum Calibration Error (MCE):-}}
The  Maximum Calibration Error (MCE) shows the the worst-case CE deviation in a bin as: 
\begin{equation}
    MCE = \max_{m \in \{1\dots M\}} |acc(B_m)-conf(B_m)|
\end{equation}

\section{Experimental Setup}
\label{sec:pagestyle}

We use the LibriSpeech (LS) dataset~\cite{librispeech} as our source domain and the GigaSpeech (GS) YouTube dataset~\cite{gigaspeech} as our target domain. Note that LS is a read speech and GS YouTube is a spontaneous speech. Hence, the domain mismatch is severe and is a hard combination for the study. Following the original DUST paper setup, we use 100 hrs of source domain data (train-clean-100) to train our teacher model~\cite{dust}. We perform all our PL filtering experiments on the GigaSpeech YouTube dev set (GS\_youtube) as an unlabeled target domain. 
Note that the difference in data distribution between source and target domains can be attributed to various factors such as different environments, speakers, channels, accents, gender, age group, topics (e.g., medical, sports, etc.), and many more. We use \textit{domain} as a general term comprising all/any of such factors. 

We use standard ESPNet~\cite{watanabe2018espnet} transformer recipes and the configurations to train our teacher model.
We use a 12 layer transformer with CTC loss for training. 
 The final model is obtained by averaging the 10 best models with the least loss on the source domain validation set. A beam width of 20 is used for inference. Following the DUST paper, we keep $T=3$ for all experiments.

  Using target domain Language Models (LMs) might hide the cause of errors. Hence, to study the performance in a transparent way, we do not use source or target domain LMs. Note that this is also helpful when target domain text is not available for training LMs.
Further, unlike multiple data augmentation schemes used in the previous DUST paper~\cite{dust} that makes the teacher model robust to the target domain, for our first three sets of experiments, we keep data augmentation minimal with standard SpecAugment~\cite{specaug} (specaug). This minimal setup helps to understand the filtering algorithm in a transparent way.

\begin{figure}
    \centering
\frame{\includegraphics[width=0.25\textwidth]{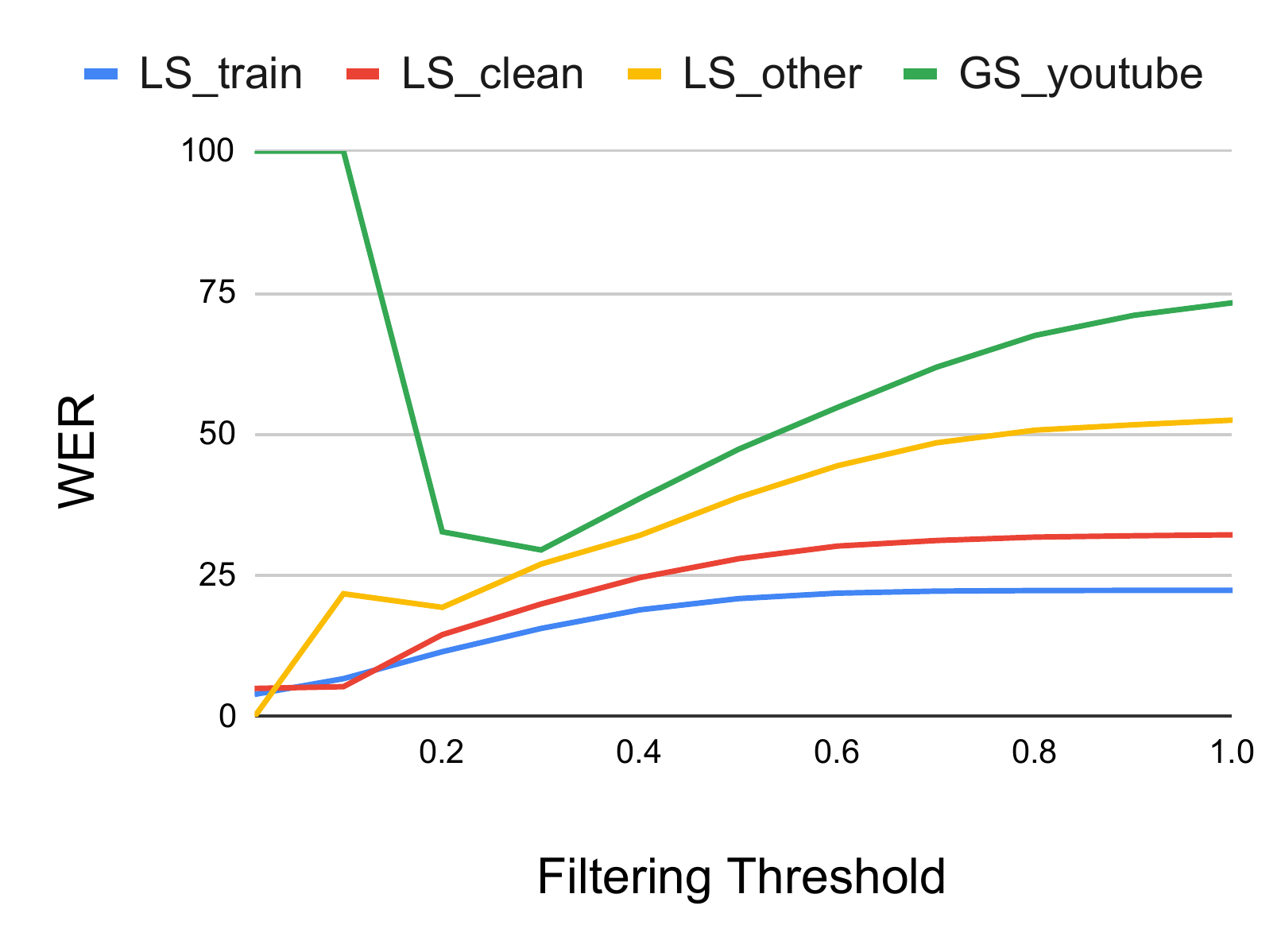}}
    \caption{DUST pseudo-label filtering: (i) \textit{Small domain mismatch:} Expected PL filtering trend on LS\_clean. (ii) \textit{Medium domain mismatch:} Small deflection on LS\_other. (iii) \textit{Severe domain mismatch:} Fails when domain mismatch is severe on GS\_youtube.}
    \label{fig:breakdust}
\end{figure}

\section{Results}
\label{sec:results}

In this section, we show the breaking point of the DUST PL filtering algorithm, demonstrate the effectiveness of various approaches that can mitigate this issue, and show how the DUST PL filtering is highly correlated with the model's calibration.

\vertspace\paragraph*{DUST PL filtering breakpoint:} 
Fig.~\ref{fig:breakdust} shows the WER for different filtering thresholds for three different variations in data distributions compared to the source domain train data (LS\_train) distribution. (i) \textit{Small domain mismatch:} It can be seen that the LS\_clean (test-clean in LS), which has data distribution close to that of source domain train set, shows expected trend. The WER increases with the threshold, while the cleaner PLs are accepted first. (ii) \textit{Medium domain mismatch:}  When we filter the PLs in the LS\_other set (dev-other in LS), which is noisy compared to LS\_clean, initially there is a small deflection in the curve at filtering threshold 0.1 to 0.2 due to slight domain mismatch. (iii) \textit{Severe domain mismatch:} When the domain mismatch is severe (GS\_youtube), DUST initially accepts PLs with very high WER, which is undesirable. This is because these utterances show high consensus (incorrectly) among their reference and three sampled predictions, but are hypothesized incorrectly (reference as well as sampled predictions).

We notice this behavior for the initial 1-2\% of total unlabelled target domain data. It is worth highlighting that this, in practice, is a significant number, as the PL filtering algorithm is used to generate PLs at a large scale. It is crucial to select good quality initial PLs precisely. For example, if there are 1000 hours in the unlabelled target domain set and the DUST PL filtering algorithm with a fixed threshold (say 0.4) filters out 100 hours to be used for training a student model. Note that 10-20 hours (out of the selected 100 hours) of utterances are wrongly filtered having noisy PLs. This is a significant number of noisy PLs in the training data, which might affect the student model training adversely. As this work focuses on PL filtering, self-training experiments are out of the scope of this paper and we leave investigations in this direction for future work.
From here on, we demonstrate all PL filtering results on the target domain GS\_youtube dataset.

\begin{figure}[h]
    \centering
\frame{\includegraphics[width=0.27\textwidth]{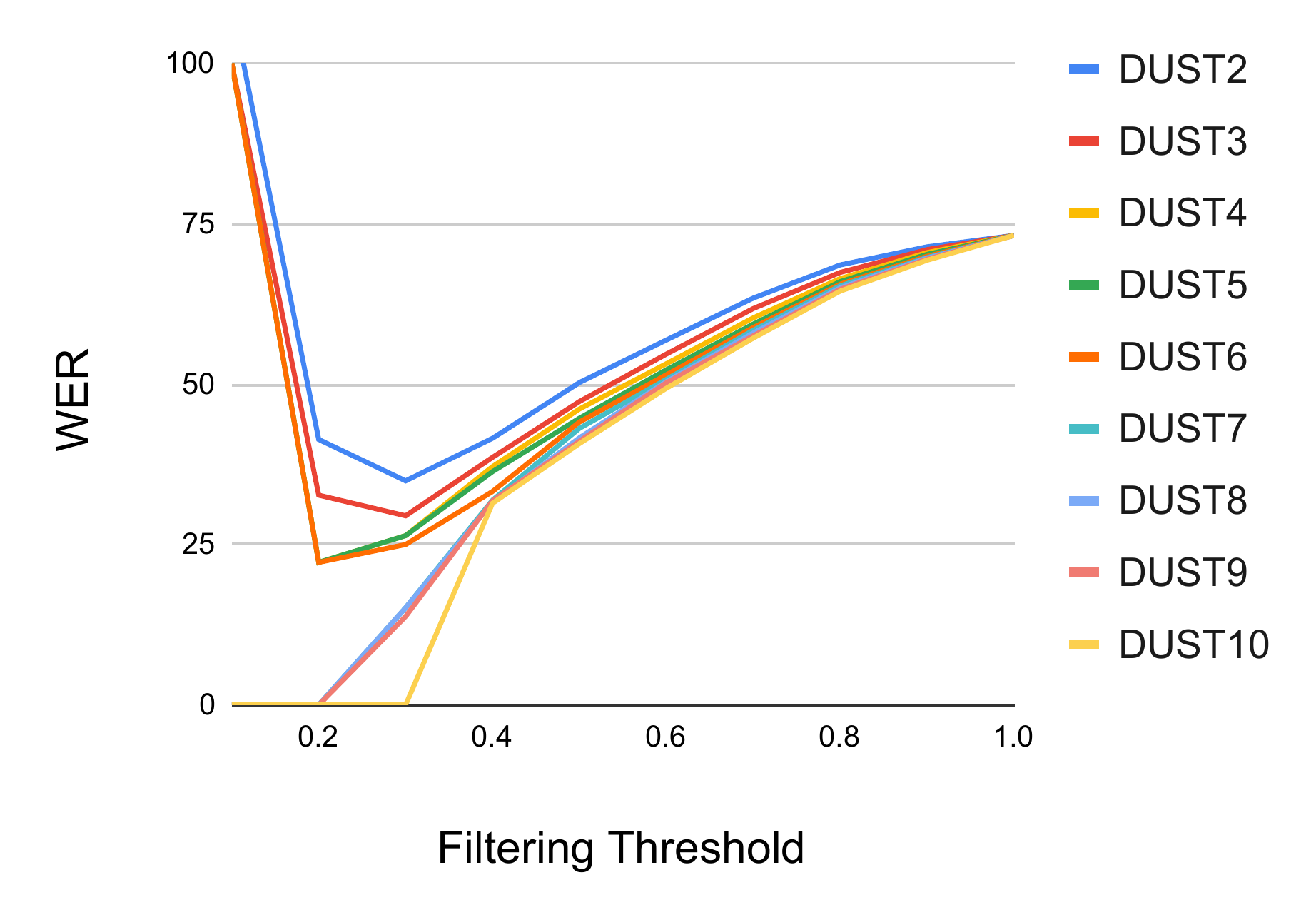}}
    \caption{\textbf{Approach-1}:  Increasing number of sampled predictions ($T$) makes PL filtering decisions stricter and eliminates bad PLs. }
    \label{fig:approach1}
\end{figure}

\vspace{-0.2cm}
\vertspace\paragraph*{Approach-1, Stricter consensus:} 
Fig.~\ref{fig:approach1} shows the plot for filtering threshold and the WER for different numbers of sampled predictions. We vary the value of $T$ from 2 to 10. It can be seen that increasing the number of sampled predictions is beneficial. It makes the filtering condition stricter as more number of samples are involved in mutual consensus leading to robust filtering decisions. With $T>6$, it eliminates the initial PLs with very high WER. 
\begin{figure}[h]
     \centering
     \begin{subfigure}[b]{0.23\textwidth}
         \centering
         \frame{\includegraphics[width=0.9\textwidth]{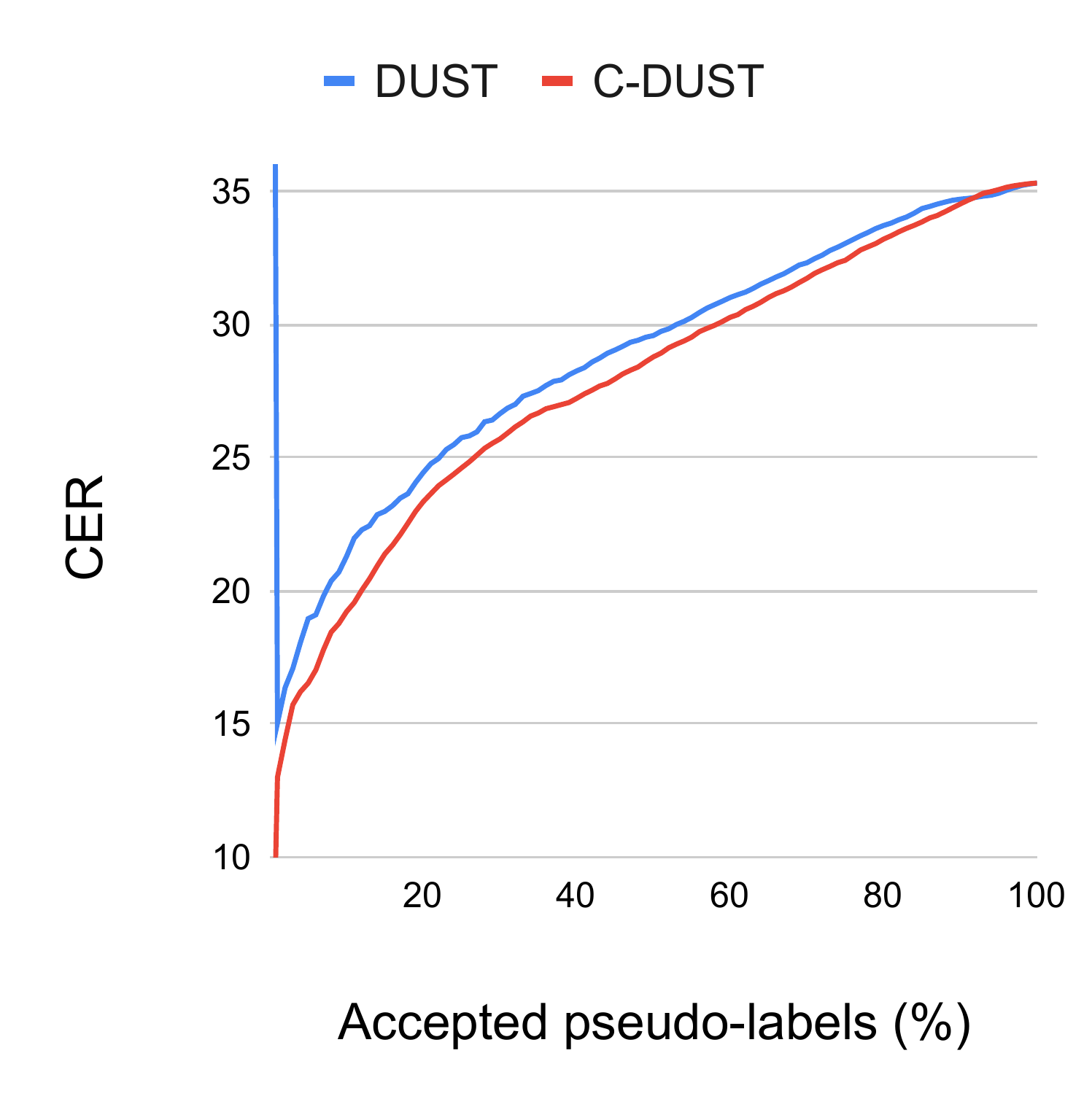}}
         \caption{T=3}
         \label{fig:filt_3}
     \end{subfigure}
    ~
     \begin{subfigure}[b]{0.23\textwidth}
         \centering
         \frame{\includegraphics[width=0.9\textwidth]{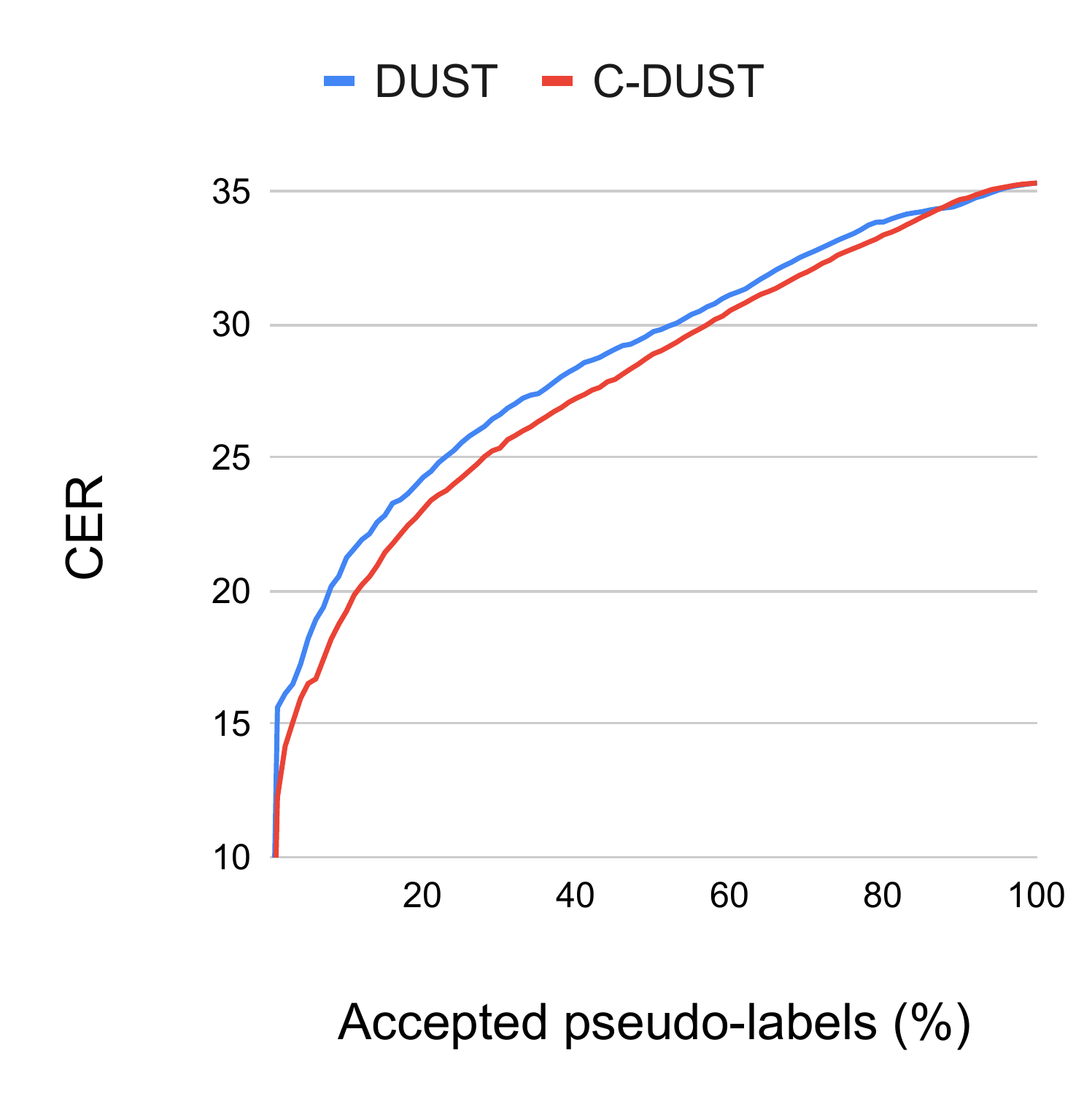}}
         \caption{T=10}
         \label{fig:filt_10}
     \end{subfigure}
        \caption{\textbf{Approach-2:} Smaller tokens (character). C-DUST filters better quality PLs compared to DUST.}
        \label{fig:dust_cdust_filtering}
\end{figure}

\vspace{-0.2cm}
\vertspace\paragraph*{Approach-2, Smaller tokens:} 
Fig.~\ref{fig:dust_cdust_filtering} shows the filtering mechanism when word-based edit distance is used as predictive uncertainty (in case of DUST) and when character-based edit distance is used as predictive uncertainty (in case of C-DUST). Note that since the range of edit distances (i.e., $pred\_uncert(x_u)$ values) in DUST and C-DUST are quite different, it is not possible to compare DUST and C-DUST for a given filtering threshold. Hence, we compare them using the percentage of accepted PLs (x-axis). Also, we notice that the WER for a given percentage of accepted PLs for both algorithms is similar, while the C-DUST shows improvement in terms of CER. This is because C-DUST precisely selects PLs based on finer units (i.e., characters) as opposed to coarse word-based DUST processing. Note that the set of utterances selected by both algorithms can be different.
Importantly, the C-DUST consistently selects better PLs compared to DUST irrespective of the number of sampled predictions ($T$). 

The improvements for C-DUST in the initial filtered PLs are more compared to those selected later. This is desirable as the initial set of filtered PLs included in the selected PL list are less noisy. Further, it is worth mentioning that as C-DUST focus on finer tokens, it is most beneficial when target domain text is not available to train LM. For example, it is very difficult to obtain specific target medical domain transcriptions to train LMs~\cite{alhanai_thesis}.

\begin{figure}[t]
    \centering
\frame{\includegraphics[width=0.22\textwidth]{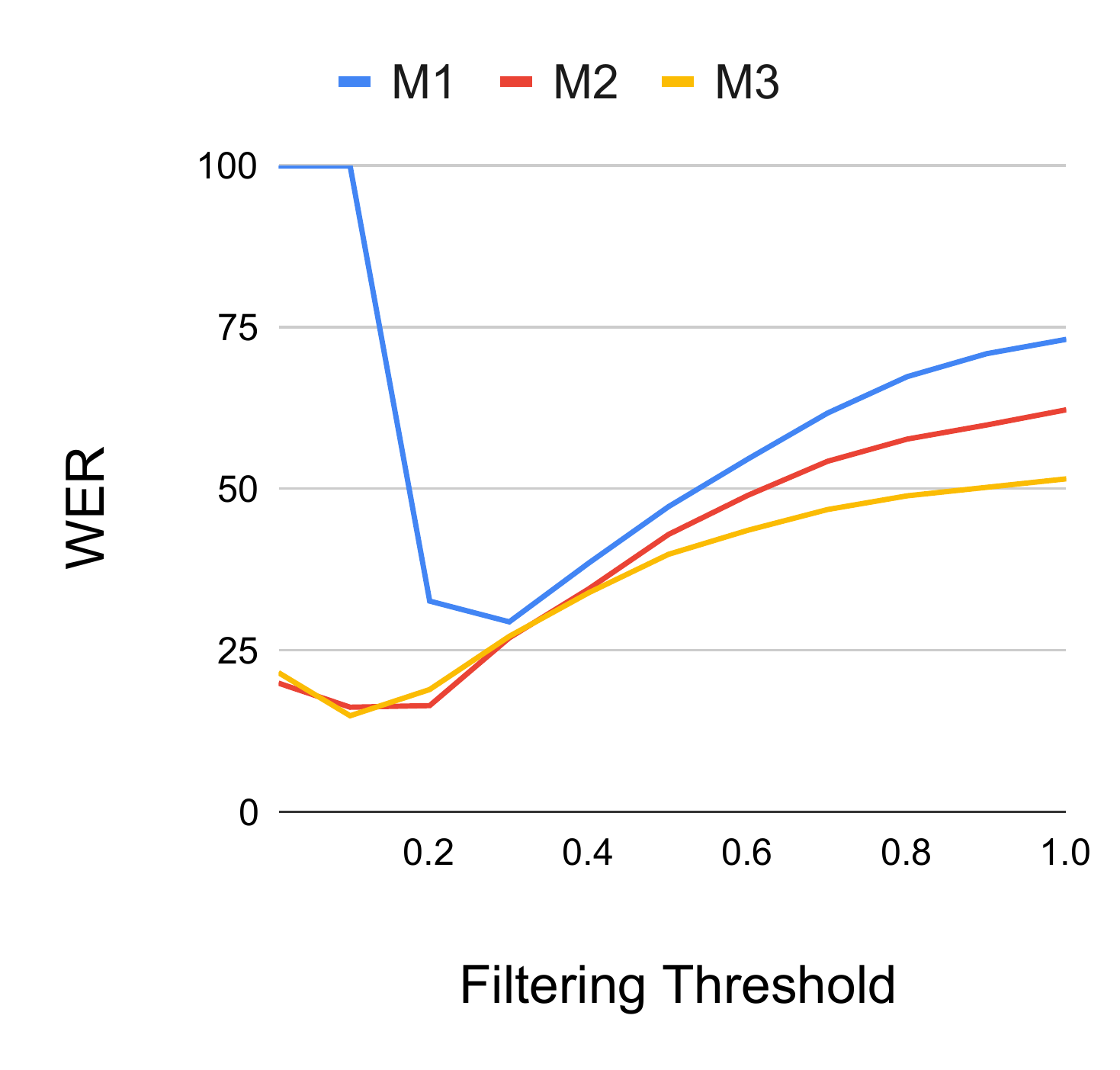}}
    \caption{\textbf{Approach-3}: Robust teacher (M2) and \textbf{Approach-4}: Robust + Diverse source data (M3). PL filtering improves from M1 to M3.}
    \label{fig:robust_largedata}
\end{figure}

\begin{table}[]
\centering
\caption{Different configurations for teacher models.}
\vspace{-0.25cm}
\resizebox{0.28\textwidth}{!}{
\begin{tabular}{@{}lll@{}}
\toprule
Models & Source domain & Data Augmentation     \\ \midrule
M1     & 100 hrs                & specaug          \\
M2     & 100 hrs                & specaug+sp       \\
M3     & 960 hrs                & specaug+sp       \\ \bottomrule
\end{tabular}}
\label{tab:modelconfigs}
\end{table}

\vertspace \paragraph*{Approach-3 and 4, Teacher model:}
 In this set of experiments, we train three teacher models covering different variations in the source domain as shown in Table~\ref{tab:modelconfigs}. For data augmentation with speed perturbation (sp), we use sp factors of 0.9, 1.0, and 1.1. Fig.~\ref{fig:robust_largedata} shows the performance of DUST PL filtering using different teacher models. It can be seen that the model M2 trained with source data augmentation using speed perturbation filters better target domain PLs compared to M1. This shows that training a robust teacher model helps in obtaining better PLs.  We use model M3 which is trained on 960 hours of labeled source domain data along with data augmentation using speech perturbation. It covers diverse conditions in the source domain. It can be seen that M3 outperforms the other two models. Hence, when source domain labeled data is available in abundance with diverse conditions, it can be used to train teacher models.
 
 It can also be seen that as we improve the teacher model from M1 to M3, the severity of selecting an initial set of bad PLs is reduced. This is because the teacher model becomes more robust and, overall the filtering process becomes more confident about PL filtering decisions (as discussed in the next section). Hence, improving the PL filtering model's confidence along with the accuracy helps alleviate the DUST filtering issues under severe domain mismatch.

\begin{figure}[]
     \centering
     \begin{subfigure}[b]{0.155\textwidth}
         \centering
         \includegraphics[width=\textwidth]{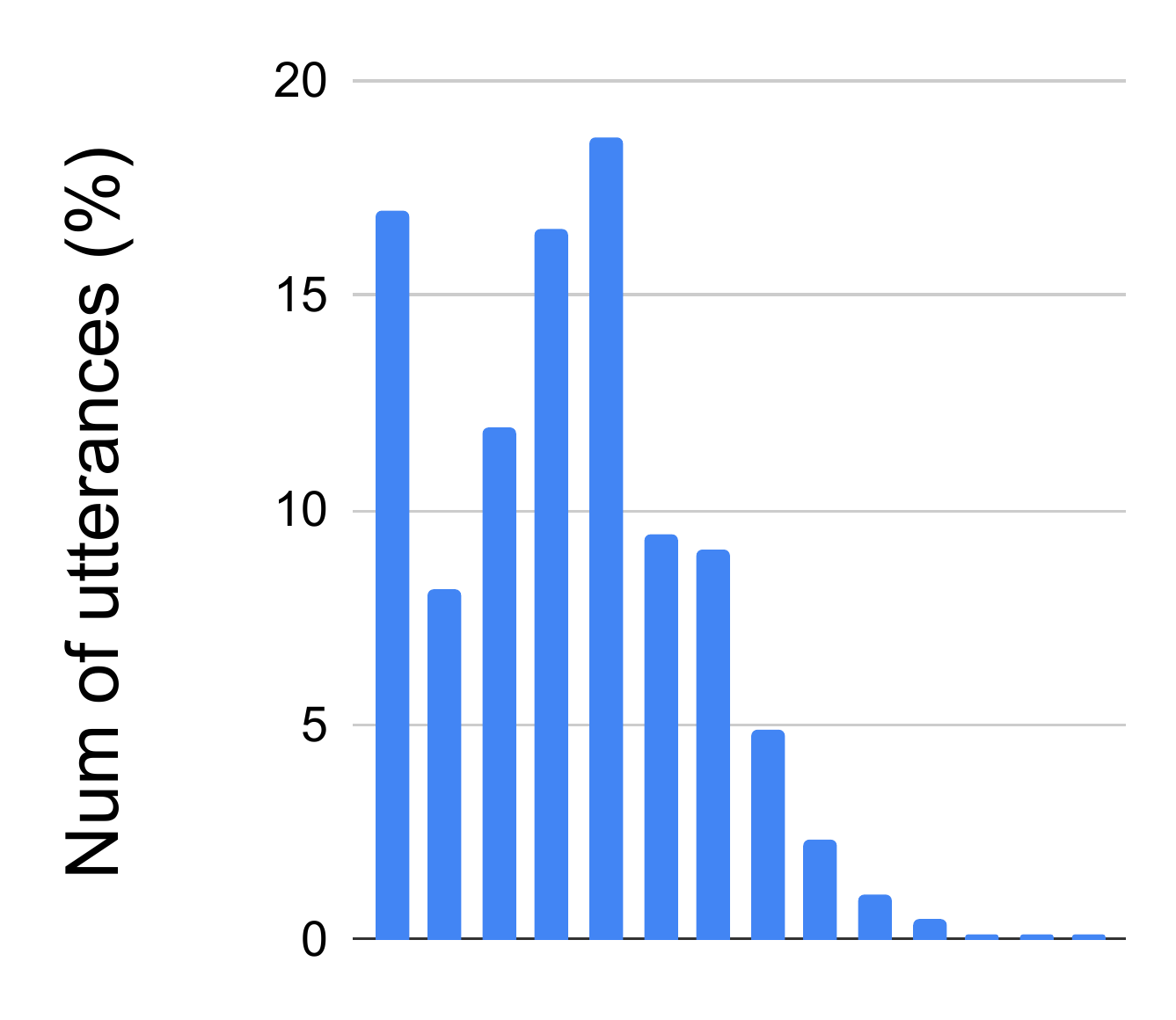}
     \end{subfigure}
     \begin{subfigure}[b]{0.155\textwidth}
         \centering
         \includegraphics[width=\textwidth]{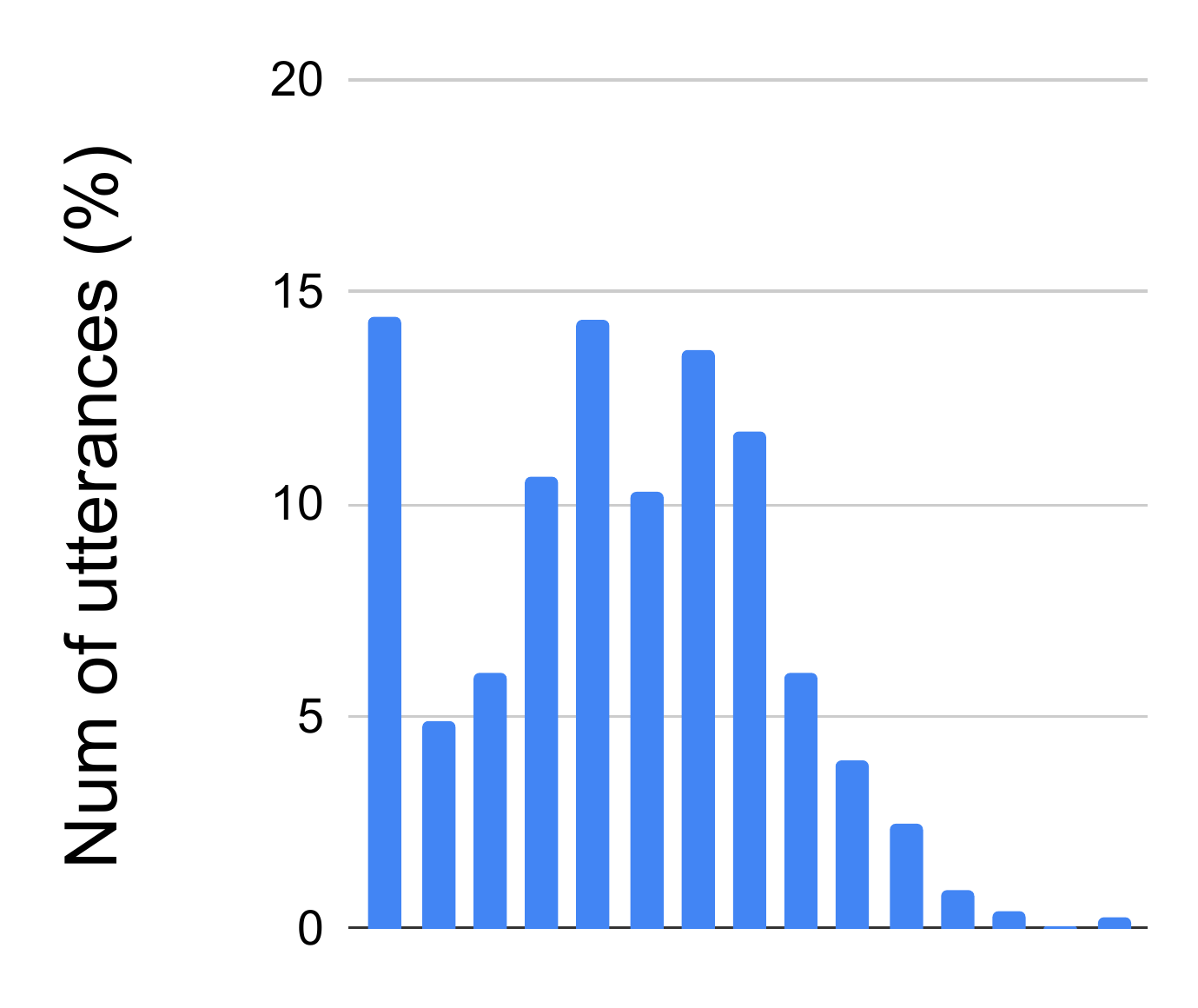}
     \end{subfigure}
     \begin{subfigure}[b]{0.155\textwidth}
         \centering
         \includegraphics[width=\textwidth]{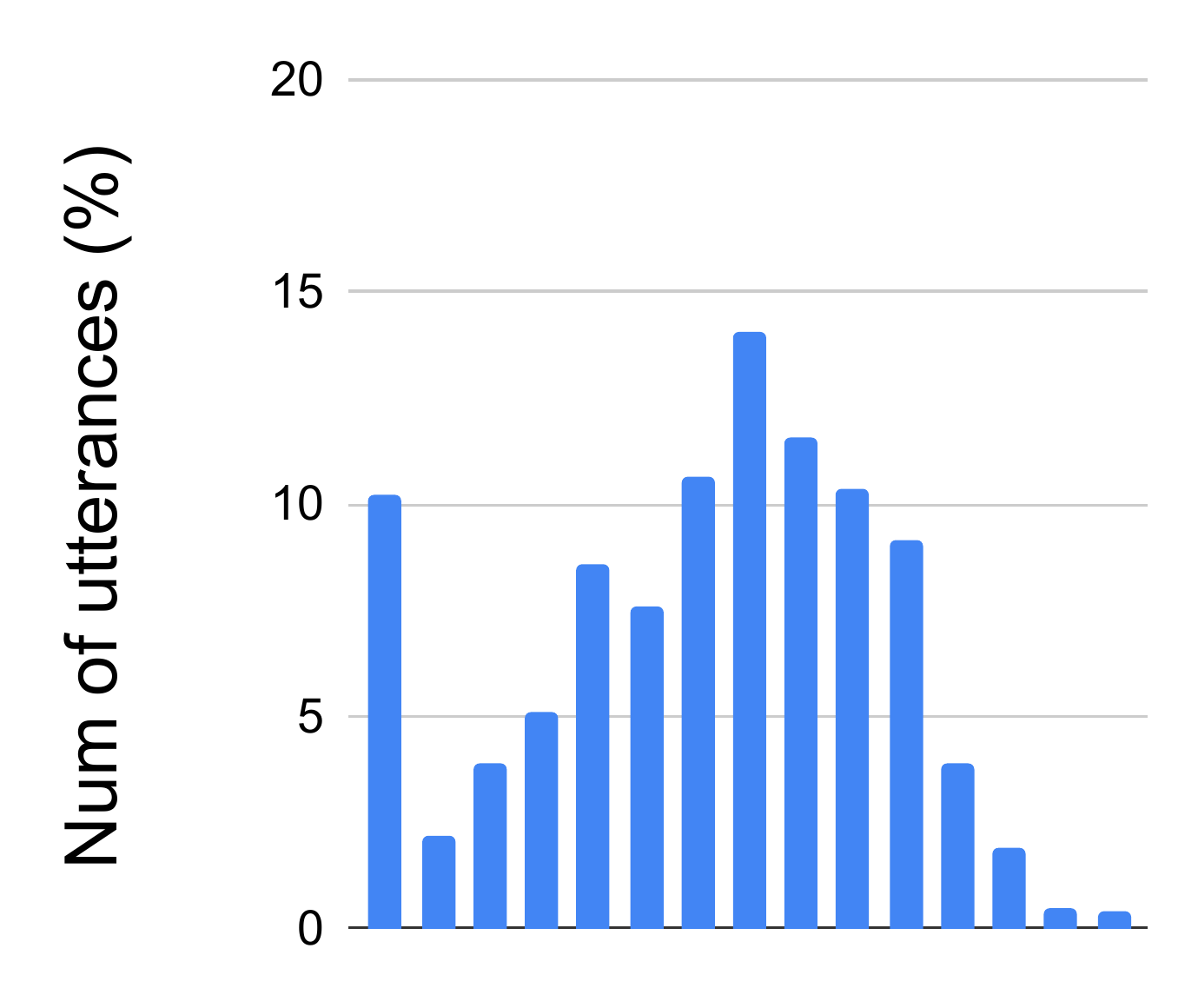}
     \end{subfigure}

     \begin{subfigure}[b]{0.155\textwidth}
         \centering
         \includegraphics[width=\textwidth]{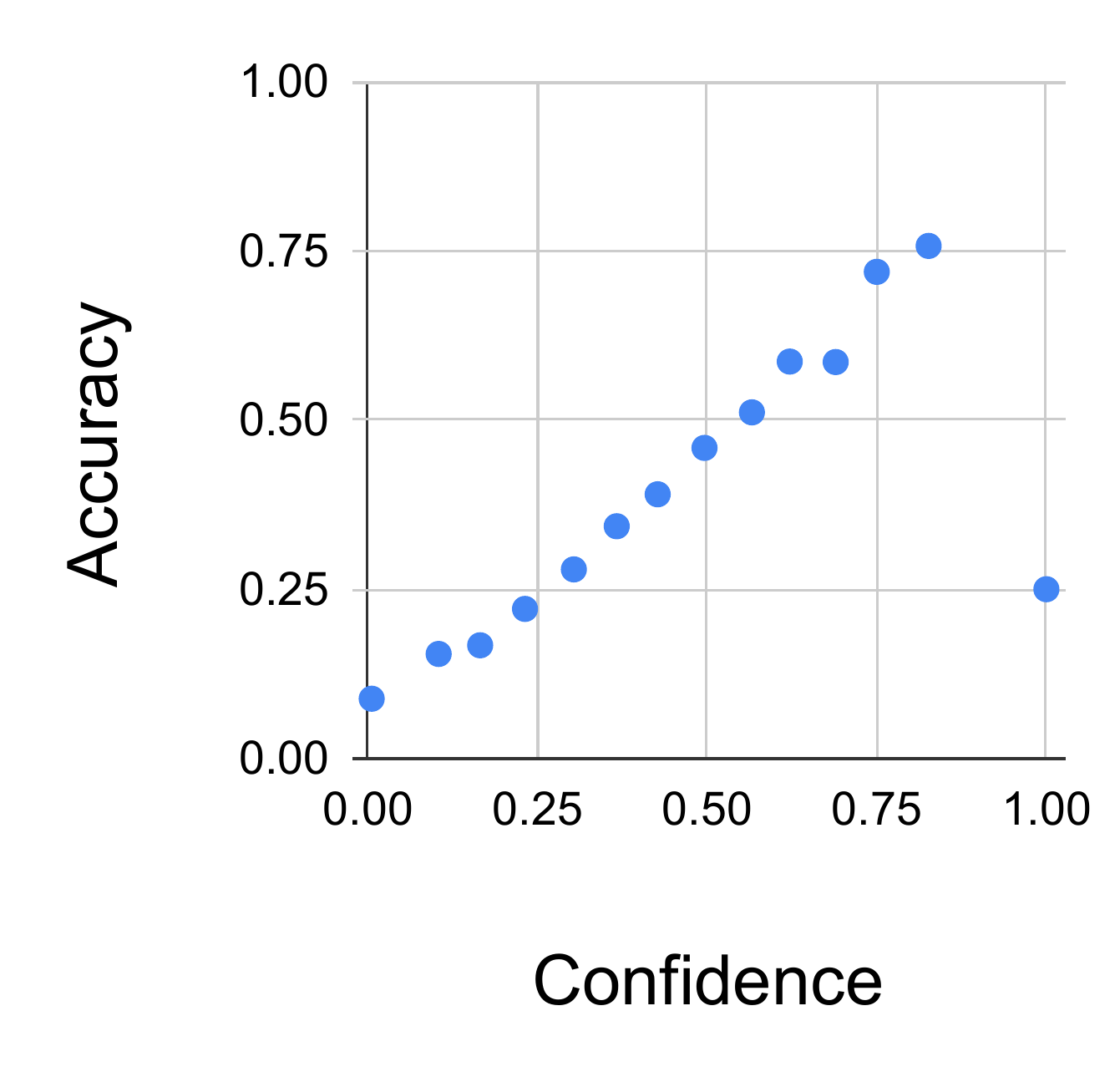}
         \caption{Teacher model M1}
         \label{fig:reliabilty:M1}
     \end{subfigure}
     \begin{subfigure}[b]{0.155\textwidth}
         \centering
         \includegraphics[width=\textwidth]{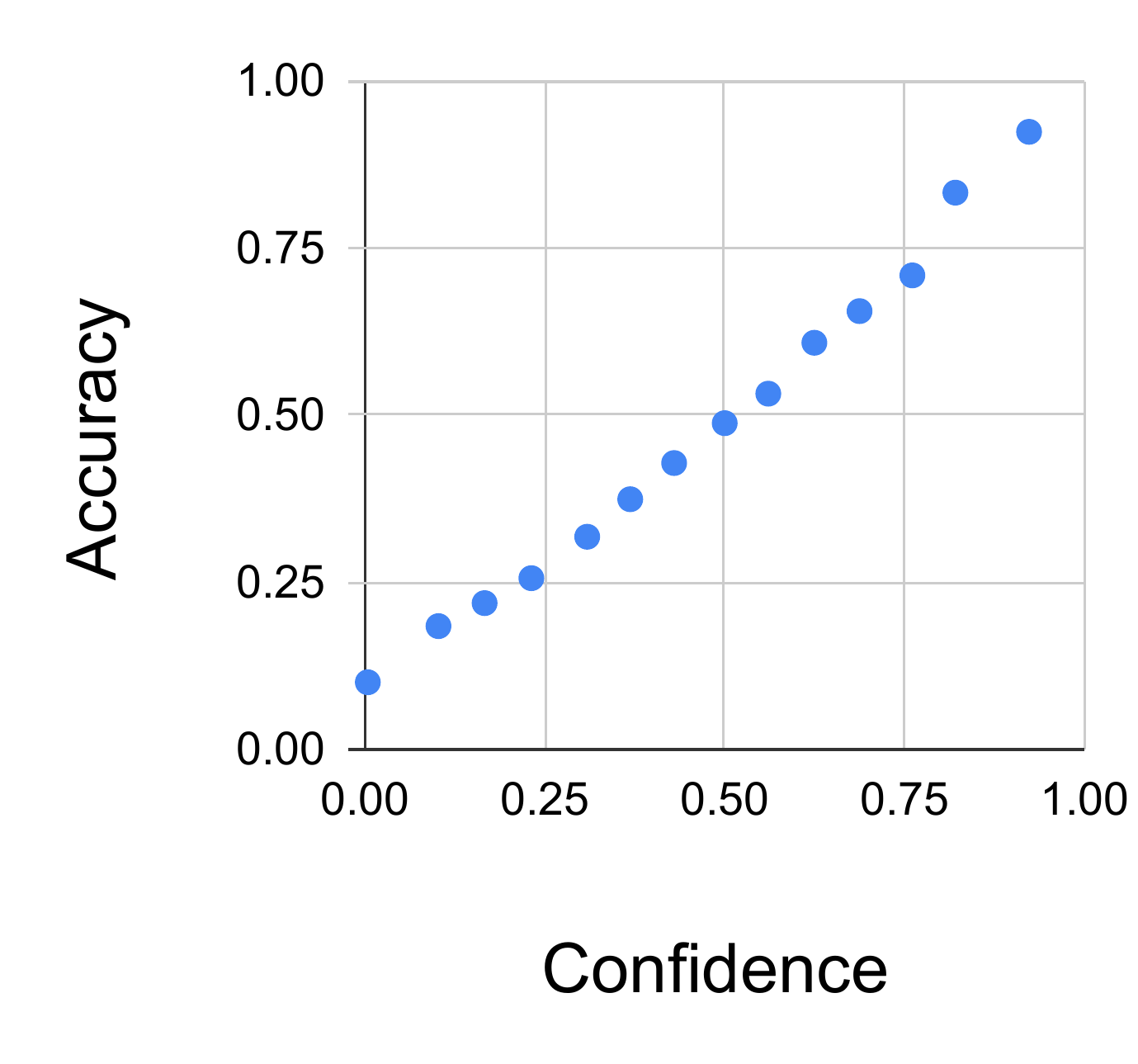}
         \caption{Teacher model M2}
         \label{fig:reliabilty:M2}
     \end{subfigure}
     \begin{subfigure}[b]{0.155\textwidth}
         \centering
         \includegraphics[width=\textwidth]{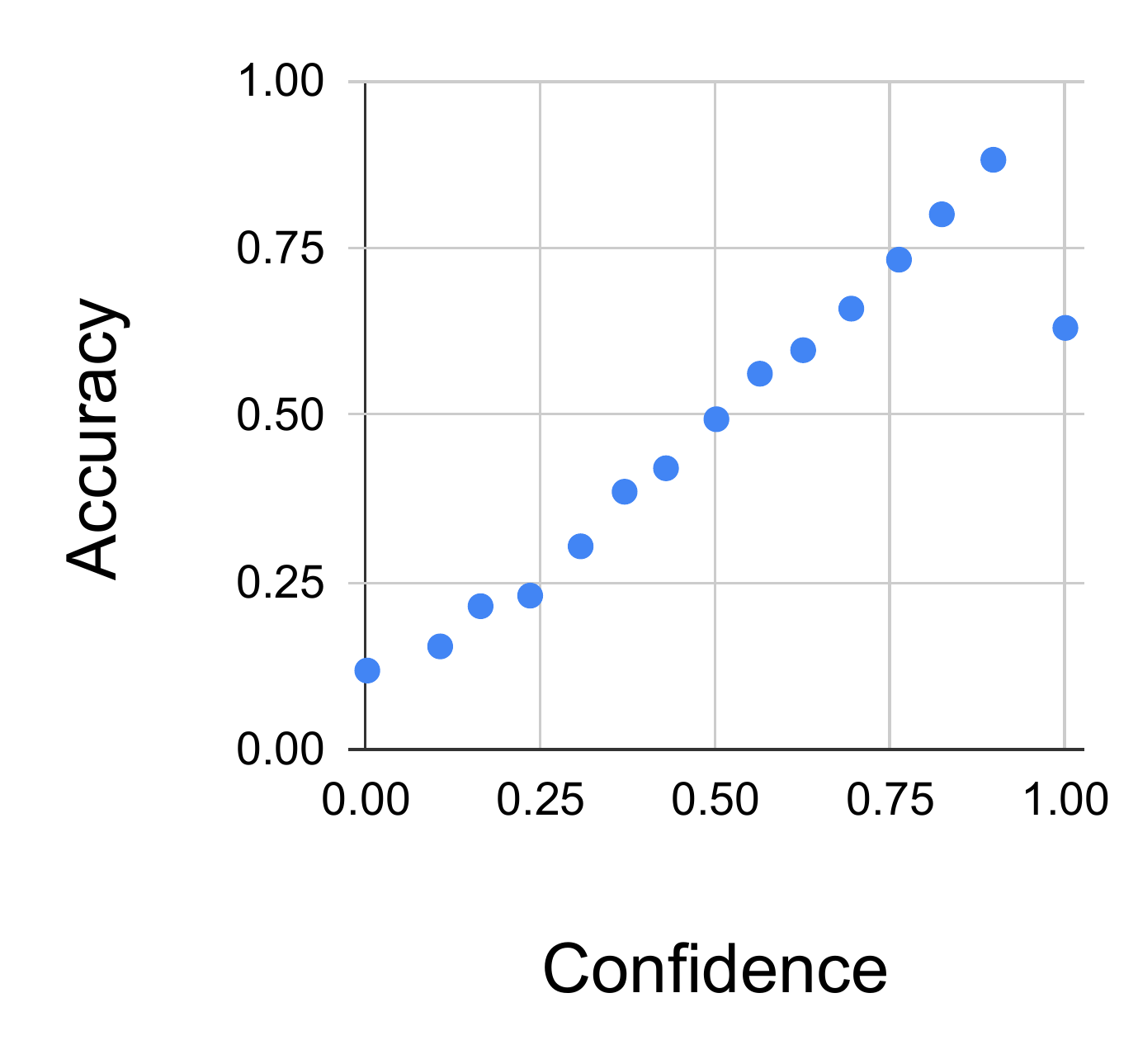}
         \caption{Teacher model M3}
         \label{fig:reliabilty:M3}
     \end{subfigure}
    
        \caption{Reliability diagram for different teacher models using DUST. Model's confidence and accuracy improve from M1 to M3.}
        \label{fig:reliabilty}
\end{figure}

\vertspace\paragraph*{Model calibration and PL filtering:}
Following~\cite{guo_calib:2017}, we plot the reliability diagram for different teacher models with DUST PL filtering in Fig.~\ref{fig:reliabilty}.
The bottom row shows confidence vs accuracy for each of $M$ bins (each dot represents a bin, $M$=15). All the bins are sorted in ascending order of bin confidence, i.e., $conf(B_m)$. A perfectly calibrated system have $acc(B_m)=conf(B_m), \forall m\in\{1,\dots,M\}$, i.e., accuracy for each bin equals to confidence. It can be seen that the trend improves as we go from M1 to M3.
The top row shows the percentage of utterances in corresponding bins. It can be seen that the M1 model's first bin ($B_1$) have more than 15\% of utterance's PLs with low confidence. As we go from M1 to M3, this distribution shifts towards the right (high confidence), and DUST PL filtering is more confident about filtered PLs quality. This strongly correlates with Fig.~\ref{fig:robust_largedata}, where filtering of PLs improves with various improvements to the teacher model (from M1 to M3).

Table~\ref{tab:calib_dust} shows the calibration errors for different teacher models with DUST PL filtering. It can be seen that the main CE metrics ECE and RCE improve (reduces) consistently from models M1 to M3.  We see a minor increase in MCE (worst-case bin error) for M3 compared to M2. This might be due to an outlier bin error which can also be observed in Fig.~\ref{fig:reliabilty:M3}. Notice that the confidence and accuracy averaged over all PLs increases from M1 to M3. This also strongly correlates with better filtering and better WER in Fig.~\ref{fig:robust_largedata}. Hence, better calibrated models with high accuracy and high confidence improve PL filtering. This finding opens up the potential future research directions in training well-calibrated PL filtering models using model calibration techniques, e.g., temperature scaling~\cite{guo_calib:2017}.

\begin{table}[]
\centering
\caption{Model calibration errors (ECE, MCE, RCE), average confidence (CNF) and average accuracy (ACC) using different teacher models with DUST PL filtering.}
\vspace{-0.25cm}
\resizebox{0.35\textwidth}{!}{
\begin{tabular}{@{}cccccc@{}}
\toprule
Models & ECE                  & MCE                  & RCE                  & CNF                  & ACC                  \\ \midrule

M1     & 0.0353               & 0.7500               & 0.0509               & 0.2485               & 0.2500               \\
M2     & 0.0318               & 0.3519               & 0.0488               & 0.3287               & 0.3477               \\
M3     & 0.0280               & 0.3700               & 0.0486               & 0.4381               & 0.4422               \\
\bottomrule
\end{tabular}}
\label{tab:calib_dust}
\end{table}

\vertspace
\section{Conclusions}
\vspace{-0.2cm}
We revisited the DUST PL filtering algorithm and found that it fails to filter correctly the PLs when domain mismatch is severe. We suggest the following approaches to eliminate or alleviate this limitation, (i) Stricter consensus, (ii) Use of smaller tokens for estimating uncertainty, (iii) Robust teacher model, and (iv) Diverse source domain data for training teacher model. We demonstrate the effectiveness of each of these approaches using YouTube speech as our target domain. Finally, we show that the quality of filtered PLs is strongly related to the model's calibration error. In the future, we plan to explore this further for training well-calibrated models for PL filtering.




\bibliographystyle{IEEEbib}

\bibliography{bib_domain_adapt}

\end{document}